\documentclass[varenna]{cimento}

\usepackage{graphicx}
\usepackage[version=3]{mhchem}   	
\usepackage{sistyle}			

\title{Sympathetic cooling of \ce{OH-} ions using ultracold Rb atoms in a dark SPOT}
\author{H. L\'opez, B. H\"oltkemeier, J. Gl\"assel, P. Weckesser \atque M. Weidem\"uller}
\institute{Physikalisches Institut, Universit\"at Heidelberg\\ INF 226, Heidelberg 69121, Germany}
\author{T. Best, E. Endres \atque  R. Wester}
\institute{Institut f\"ur Ionenphysik und Angewandte Physik, Universit\"at Innsbruck\\ Technikerstr. 25/3, 6020 Innsbruck, Austria}
\shortauthor{H. L\'opez et al.}

\begin{document}

\maketitle

\begin{abstract}
We are developing a new hybrid atom-ion trap to study the interaction of ultracold rubidium atoms with mass-selected \ce{OH-} molecules. 
The ions are trapped inside an octupole rf-trap made of thin wires instead of the commonly used rods. 
This ensures good optical access to the center of the trap where the ions can be overlapped with laser cooled rubidium atoms stored in a dark spontaneous force optical trap (dark SPOT).
This setup provides high collision rates since the density in a dark SPOT is about one order of magnitude higher than in a standard magneto-optical trap. 
Further, inelastic collisions with excited atoms are suppressed since almost all atoms are in the ground state.
Numerical simulations of our setup using SIMION predict that cooling of the ions is feasible. \footnotemark
\end{abstract}

\footnotetext{To appear in the Proceedings of the International School of Physics Enrico Fermi, Course 189 "Ion Traps for Tomorrow's Applications".}

\section{Introduction}
\label{Introduction}

The study of increasingly complex quantum systems requires constant development of new techniques for their preparation and manipulation.
One of these systems that is of great interest is cold molecular ions.
Among a wide range of possible applications, they play an important role in the investigation of quantum chemistry and fundamental physics \cite{Roth2008,Carr2009}.
A particular advantage of molecular ions over neutral molecules is that they can be trapped with radio frequency traps. 

A widely used technique with the potential to cool all degrees of freedom is buffer gas cooling. 
However, using helium, the coldest commonly available buffer gas, the temperature is limited to a few Kelvin, i.\,e. two orders of magnitude warmer than the desired ultracold regime \cite{Pearson1995,Gerlich1992}.
To overcome this limit, laser cooled atomic ions have been used to sympathetically cool molecular ions \cite{Molhave2000}.
With this approach, molecular ions at translational temperature as cold as a few tens of \SI{}{mK} have been created \cite{Tong2010}.
The main drawback is that due to the long range character of the Coulomb interaction, the molecules' internal degrees of freedom cannot be cooled. 
As a result, the molecules' internal temperature is much higher than the translational temperature.

One way to reach both translational temperatures in the \SI{}{mK} regime and cool the internal degrees of freedom is to use hybrid atom ion traps (HAITrap) for laser cooled neutral atoms and molecular ions.
A recent review provides further details and applications of such traps \cite{Haerter2013}.
Most HAITraps consist of a radio frequency ion trap (rf-trap) superimposed with a magneto-optical trap (MOT), such that the ions are immersed in a cloud of laser cooled atoms \cite{Hudson2009,Smith2005}.
So far, translational cooling of molecular ions using neutral atoms has not been observed.
Nevertheless, cold atoms have been used to internally cool translationally cold molecular ions \cite{Rellegert2013}.

In contrast, for atomic ions, first experimental signatures of sympathetic cooling by neutral ultracold atoms have already been observed.
Using atomic ions of the same species as their neutral partners (e.g. \ce{Rb+ + Rb} \cite{Ravi2012}, \ce{Na+ + Na} \cite{Sivarajah2012}) sympathetic cooling was experimentally demonstrated by measuring the ions' time-of-flight distribution. 
Also using laser precooled ions, it has been shown that starting in the \SI{}{mK} regime it is feasible to subsequently cool the ions further using neutral atoms \cite{Makarov2002}.

In this proceedings article we report on the status of a new HAITrap collaboratively built by Physikalisches Institut Heidelberg and Institut f\"ur Ionenphysik und Angewandte Physik Innsbruck.
It aims to investigate the effects of sympathetic cooling and to study chemical reactions in the ultracold temperature regime.
Specifically, our new hybrid trap combines a linear octupole rf-trap \cite{Walz1994} for \ce{OH-} ions and a dark spontaneous force optical trap (dark SPOT) \cite{Ketterle1993} for rubidium atoms. 
The use of high-order ion traps is supposed to reduce rf-heating due to the micromotion in comparison to normal linear Paul traps \cite{Gerlich1992}. 

\section{Sympathetic cooling of ions using ultracold atoms}
\label{thermailzation}

Sympathetic cooling conditions in a hybrid trap strongly depend on the geometry of the trap and the choice of buffer gas \cite{Ravi2012,Wester2009,Green2007,Hudson2013}.
Unfortunately, not all collisions in an rf-trap lead to cooling of the ions.
Collisions can get the ion motion out of phase with the rf-field, effectively leading to an increase of the micromotion's energy.
This energy is then subsequently transferred to the secular motion. 
As a consequence the ions are heated up.
In general, at the outer parts of the trap where the ions' micromotion is at its maximum, collisions lead to overall heating \cite{Cetina2012}.
At the trap center where the ions' motion is dominated by the secular motion, collisions lead to cooling.
The largest possible energy transfer per collision is achieved using atoms with the same mass as the ions.
However, the heavier the atom the bigger is the heating effect at the outer parts of the trap since the average phase shift is increased.
Therefore, in a linear Paul trap homogeneously filled with buffer gas, cooling can only be achieved if the buffer gas atoms are not much heavier than the ions \cite{Zipkes2011}.

To avoid this limitation two main improvements can be made \cite{Gerlich1992,Gerlich1995}. 
Firstly, higher order traps can be used.
They provide a large field-free center with a steep potential barrier towards the edges.
Hence, on average the ions spend more time in the field free center where collisions lead to cooling.
Secondly the cooling agent can be confined to the field free center of the trap reducing the heating effects even further.
As compared to standard cryogenic buffer gas, which at best can be collimated to a beam, the use of hybrid traps makes it possible to confine laser cooled atoms to any desired region inside the trap. 
In this configuration, using heavy atoms no longer precludes cooling of the ions.

Nevertheless, how effectively the ions can be cooled not only depends on the sympathetic cooling rate, but several heating and loss mechanisms have to be considered.
Direct loss channels are inelastic collisions with excited atoms and chemical reactions which change the charge state of the ions.
Ion heating is caused by collisions with background gas which can lead to a rethermalization to room temperature.
Furthermore, imperfections in the rf-traps electrodes and surface charge effects perturb the ions' trajectory leading to increased rf-heating.
In order to cool the ions, the cooling rate has to exceed the combined heating rates.

\section{Hybrid atom-ion trap}

In our setup we combine a dark SPOT with an octupole linear rf-trap. 
With this HAITrap we want to investigate the possibility to sympathetically cool molecular ions using rubidium atoms.

There are two main advantages of using a dark SPOT instead of a conventional MOT. 
Firstly, in a dark SPOT the fraction of atoms in the excited state is highly reduced. 
This way the inelastic collision rate of excited atoms with ions is very low increasing the ions' lifetime in the trap. 
Secondly, the atom density in a dark SPOT exceeds that of a MOT by about one order of magnitude leading to enhanced cooling rates. 
Therefore, using a dark SPOT, the large field-free region of the octupole trap is used most efficiently, ensuring large collision rates and long ion lifetimes in the trap.

The first system we want to investigate is \ce{OH- + Rb}, which is expected to show very large internal de-excitation rates \cite{Gonzales-Sanches2008}.
As reported in ref. \cite{Deiglmayr2012}, this system has two loss channels, one of which can be suppressed. 
The first loss channel is the inelastic collision of one excited rubidium atom with a hydroxide ion, 
\begin{equation}
\ce{OH- + Rb$^*$ -> OH- + Rb + E_{kin}}, 
\label{eq:inel}
\end{equation}
where the excitation energy of the rubidium atoms (about 1.6 eV) is transferred as kinetic energy to the ion leading to its removal from the trap. 
By using a dark SPOT, this loss channel can be mitigated since only less than 5\% of the rubidium atoms are in the excited state. 

The other loss channel is associative detachment (AD): 
\begin{equation}
\label{eq:ad}
\ce{OH- + Rb -> RbOH + e-}.  
\end{equation}
Here, the neutral rubidium atoms associate with the hydroxide and thereby eject an electron out of the ion. 
Rubidium hydroxide is formed, which as a neutral molecule no longer interacts with the ion trap. 
The main contribution to the loss rates observed in ref. \cite{Deiglmayr2012} is attributed to AD. 
However, recent numerical calculations of associative detachment pathways for this system suggest that this dissociation channel is energetically inaccessible for the vibrational ground state, which is already reached at room temperature \cite{Byrd2013}. 
Therefore, the significance of this channel has to be investigated in more detail with our new setup. 

\begin{figure}
\begin{center}
\includegraphics[width=0.75 \columnwidth]{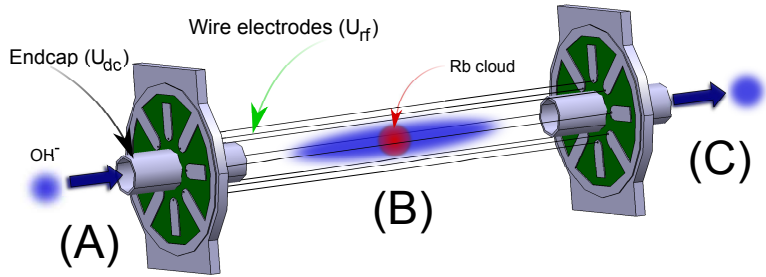}%
\end{center}
\caption{Schematics of the HAItrap. 
The octupole rf-trap has a diameter of \SI{6}{mm} and is \SI{34}{mm} long. 
(A) The ion trap is loaded from a plasma-discharge ion source through the hollow end cap. 
The \ce{OH-} ions are mass-selected by time of flight. 
(B) The ions are trapped and thermalize with He-buffer gas at \SI{293}{K}. 
The Rb dark SPOT is loaded. 
(C) After a variable interaction time, the ions are extracted and detected by a time of flight mass-spectrometer.}%
\label{experiment}
\end{figure}

Figure \ref{experiment} illustrates schematically our hybrid trap.
First, the ions are created in a plasma discharge ion source, from where they are guided down a \SI{80}{cm} long drift region for mass selection before entering the ion trap.
Once the ions are trapped, the dark SPOT is turned on such that the ions are immersed into the cloud of cold atoms.
After a variable interaction time, the ions are extracted from the trap, pass another drift region and are then detected by a microchannel plate detector (MCP).
In order to measure the ions' temperature either the time of flight distribution can be used or the ions density distribution in the trap can be directly measured performing electron detachment tomography \cite{Trippel2006}.

The dark SPOT is a widely used technique to overcome the density limitations of a standard MOT.
The atom density achievable with a MOT is mainly limited by two factors: inelastic collisions between atoms and reabsorption of scattered light of wrong polarization \cite{Anderson1994}. 
Both channels are suppressed in a dark SPOT, leading to an atom density increase of about one order of magnitude.

A MOT for rubidium atoms requires two laser frequencies to trap the atoms.
The $5^2S_{1/2} \ (F=3) \rightarrow 5^2P_{3/2} \ (F'=4)$ transition is a closed cooling cycle. 
Unfortunately, due to the small separation of the $5^2P_{3/2} \ (F'=4)$ and $(F'=3)$ states the atoms can be excited to the $(F'=3)$ state from where they can fall into the $5^2S_{1/2} \ (F=2)$ state which is a dark state for the cooling transition. 
Therefore, the $5^2S_{1/2} \ (F=2) \rightarrow 5^2P_{3/2} \ (F'=3)$ transition has to be pumped as well to repump these atoms back into the cooling cycle. 

In a dark SPOT the repumping beam is blocked at its center. 
Hence, any trapped atoms in this region fall into the dark state $5^2S_{1/2} (F=2)$. 
Most of these atoms are already cold enough to remain trapped and as soon as they reach the outer parts of the cloud, they are pumped back into the cooling cycle.
This leads to a suppression of both mentioned density limitations.
With this configuration we can trap up to $4\times10^8$ atoms with a peak density of $3\times 10^{11}$ \SI{}{atoms/cm^3} in the dark SPOT, whereas the peak density in the MOT does not exceed $3\times 10^{10}$ \SI{}{atoms/cm^3}.
Our atom cloud has a FWHM size of maximally \SI{6}{mm}. 

To achieve fast loading times of the trap we use a 2D MOT with an atomic flux of $10^9$ atoms per second and a beam divergence of less than \SI{50}{mrad}. 
The mean velocity of the precooled beam from the 2D MOT matches the trapping velocity of the dark SPOT, increasing the trapping efficiency. 
Experimental details and the characterization of this source can be found in \cite{basti2011}.

When designing an ion trap that is suited for a HAItrap, certain factors have to be considered.
Most importantly, good optical access to the trap center is needed in order to laser-cool atoms at this position.
The most commonly used ion traps consist of solid rods to which the rf-voltage is applied.
In our setup these rods have been replaced by thin wires, which leave enough space for the laser beams, but results in a less homogeneous potential.

\begin{figure}[ht]%

\begin{center}
\includegraphics[width=0.75 \columnwidth]{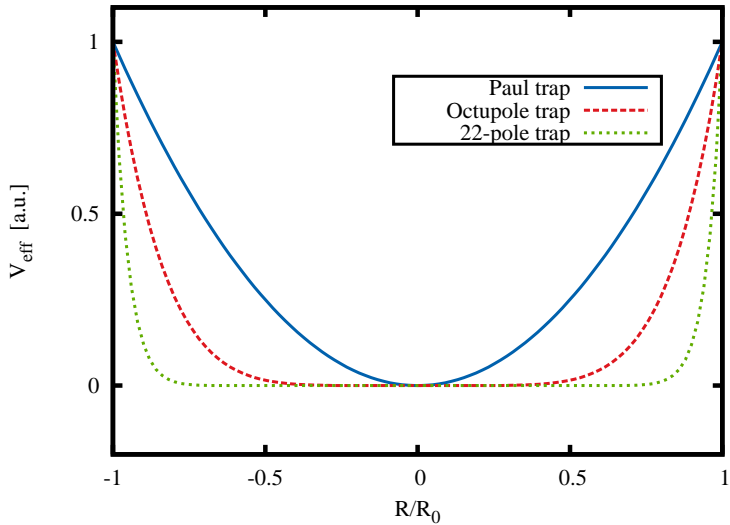}

\end{center}
\caption{Effective radial potential ($V_{eff}$) distributions for different number of poles in an rf-trap: the higher the pole number, the flatter is the effective potential at its radial center, where $R_0$ represents the position of the rf electrodes.}%
\label{Veffmultipletrap}%
\end{figure}

Another important factor is the number of wires used, which determines the shape of the trap effective potential (see fig. \ref{Veffmultipletrap}).
For the smallest possible number of electrodes, i.\,e. the classical quadrupole Paul trap, the best optical access can be obtained.
For higher order traps optical access is limited but much better cooling conditions can be reached.
Thus, the octupole trap is a good compromise between a large ion cooling rate and good optical access.  
In our setup, the octupole rf-trap has a diameter of \SI{6}{mm} (matching the size of the atom cloud) and is \SI{34}{mm} long. 
The end cap electrodes for axial confinement are hollow cylinders, which allow loading the trap from this direction.  

In order to estimate the expected cooling rates in our setup, numerical simulations using SIMION \cite{simion} have been performed.
In these simulations, the electric field of the rf-trap is solved dynamically by the Laplace equation and the ions' trajectory is calculated based on this field.
To simulate the interaction with the cold atoms, simple elastic collisions between two pointlike particles have been included into the simulation, neglecting all internal degrees of freedom.

\begin{figure}[ht]%
\begin{center}
\includegraphics[width=\columnwidth]{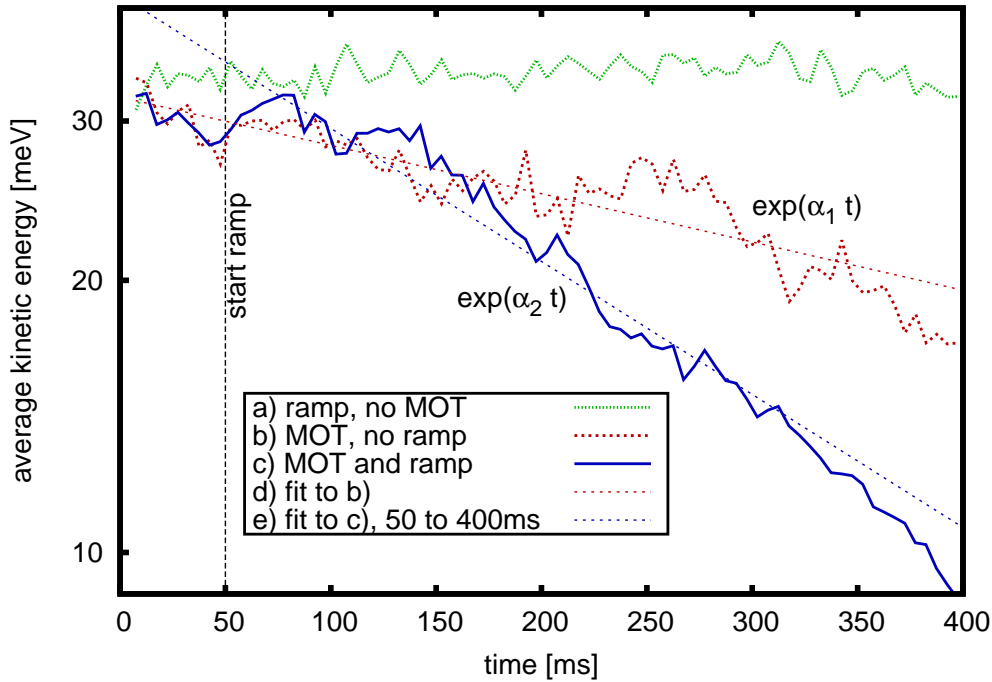}%
\end{center}
\caption{Time evolution simulations of the average ion kinetic energy in the HAItrap. a) The rf voltage is ramped down without the presence of atoms. The average kinetic energy remains the same. No cooling is observed. b) The ions interact with the atoms but no ramp is applied. Sympathetic cooling is observed. c) The ultracold atom cloud interacts with the ions while the rf voltage is ramped down. Sympathetic cooling is most efficient. d) \& e) are exponential fits to the simulation data. The cooling rate is increased from $\alpha_1 =$ \SI{1.2}{s^{-1}} to $\alpha_2 =$ \SI{3.4}{s^{-1}} by applying the ramp on the rf-voltage. e) only fits data after the rf is ramped down.} 
\label{simdensity}%
\end{figure}

All simulations have the same starting conditions: the number of simulated ions is $200$, the initial rf-voltage applied to the wires is $U_{rf} =$ \SI{300}{V} and the voltage applied to the endcaps is $U_{dc} =$ \SI{20}{V}.
All ions start at the trap center with an energy equivalent to room temperature. 
As the number of ions and therefore the ion density is small, Coulomb repulsion between the ions was neglected in the simulations.
In previous experiments we have observed that Coulomb repulsion has to be taken into account when trapping more than 500 ions at room temperature.
The simulated MOT is a Gauss-distributed atom cloud with a peak density of $4\times 10^{12}$ \SI{}{atoms/cm^3} and a FWHM size of \SI{1}{mm}. 
The chosen peak density is one order of magnitude larger than in our experiment to shorten the computation time.  
The times mentioned in the following as well as all times in fig. \ref{simdensity} are scaled to the actual experimental conditions. 
Therefor the time was scaled linearly with the MOT density, meaning that a density increase by a factor of ten is compensated by scaling up the time axis by the same factor.
On short time scales we have confirmed that this procedure does not change the result of the simulations within a ten percent margin.

Results of the simulations are shown in fig. \ref{simdensity}, where the average ion kinetic energy (in \SI{}{meV}) versus time is shown.
As pointed out in section \ref{thermailzation} sympathetic cooling of the ions is limited by collisions in the outer part of the trap due to an induced phase shift between the ions motion and the rf voltage. 
This effect can be reduced by gradually ramping down the amplitude of the rf voltage. 
In order to verify this we simulated three different scenarios. 
In fig. \ref{simdensity}b the MOT was turned on after \SI{10}{ms} and sympathetic cooling of the ions is observed. 
In fig. \ref{simdensity}c an additional ramp to the rf-voltage down to \SI{50}{V} was applied, starting at \SI{50}{ms}. 
As expected, this ramp leads to an enhancement of the cooling rate.  
Due to the long computation times, within the simulated time frame, no steady state was reached.
To show that the ramp itself does not change the ions' average kinetic energy also a simulation without a MOT but with the same ramp on the rf-voltage as in (c) was performed (\ref{simdensity}a).
In conclusion we observe sympathetic cooling by cold atoms within a time frame of a second. 
Our simulations suggest that the cooling rate can be enhanced by about a factor of three if an additional ramp to the rf-voltage is applied. 
Nevertheless, the simulation neglects additional heating effects and therefore the experimental cooling rate should be smaller than our simulations suggest.

\section{Conclusion \& outlook}
\label{conclusion}
 
We have presented the design of a new hybrid trap for neutral atoms and ions which can be used to sympathetically cool molecular ions. 
The first simulation results are presented predicting sympathetic cooling of the ions to be possible with our setup. 

In contrast to the setup used in ref. \cite{Deiglmayr2012} our design increases the atom density in the MOT by using a dark SPOT. 
This enhances the collision rate between cold atoms and ions, surpassing the cooling rate of previous experiments that used a regular MOT. 
Additionally, our design reduces competing thermalization due to collisions with the residual gas by further decreasing the vacuum pressure. 

We aim to study cold chemical reactions between molecular anions and neutral atoms.
Our setup enables us to measure reaction rates of all kinds of ions with rubidium. 
In particular, studying the interaction between neutral rubidium and hydrated water clusters \ce{OH- (H2O)_n} can give more insight into the transition from gas to condensed phase environments \cite{Xantheas1995}.
It has been shown, that such type of clusters can be trapped in multipole rf traps \cite{Greve2010}. 
Finally, we aim to develop a theory describing the achievable cooling rates in our HAITrap.

\appendix

\acknowledgments

We specially acknowledge Johannes Deiglmayr for his contribution at an earlier stage of the experiment and the financial support by the BMBF within the framework of "FAIR-SPARC" under contract number 05P12VHFA6 and the Heidelberg Center for Quantum Dynamics. B.H. acknowledges support by HGS-Hire.

\end{document}